\documentstyle[twoside,fleqn]{article}

\catcode`\@=11
\def\fileversion{v2.6}
\def\filedate{24 November 1993}

\newdimen\@bls                    
\@bls=\baselineskip               
\advance\@bls -1ex                
\newdimen\@eps                    %
\def\section{\@startsection{section}{1}{\z@}
  {1.5\@bls plus 0.5\@bls}{1\@bls}{\normalsize\bf}}
\def\subsection{\@startsection{subsection}{2}{\z@}
  {1\@bls plus 0.25\@bls}{\@eps}{\normalsize\bf}}
\def\subsubsection{\@startsection{subsubsection}{3}{\z@}
  {1\@bls plus 0.25\@bls}{\@eps}{\normalsize\bf}}
\def\paragraph{\@startsection{paragraph}{4}{\parindent}
  {1\@bls plus 0.25\@bls}{0.5em}{\normalsize\bf}}
\def\subparagraph{\@startsection{subparagraph}{4}{\parindent}
  {1\@bls plus 0.25\@bls}{0.5em}{\normalsize\bf}}

\def\@sect#1#2#3#4#5#6[#7]#8{\ifnum #2>\c@secnumdepth
  \def\@svsec{}\else
  \refstepcounter{#1}\edef\@svsec{\csname the#1\endcsname.\hskip0.5em}\fi
  \@tempskipa #5\relax
  \ifdim \@tempskipa>\z@
    \begingroup
      #6\relax
      \@hangfrom{\hskip #3\relax\@svsec}{\interlinepenalty \@M #8\par}%
    \endgroup
    \csname #1mark\endcsname{#7}\addcontentsline
      {toc}{#1}{\ifnum #2>\c@secnumdepth \else
        \protect\numberline{\csname the#1\endcsname}\fi #7}%
  \else
    \def\@svsechd{#6\hskip #3\@svsec #8\csname #1mark\endcsname
      {#7}\addcontentsline{toc}{#1}{\ifnum #2>\c@secnumdepth \else
        \protect\numberline{\csname the#1\endcsname}\fi #7}}%
  \fi \@xsect{#5}}

\long\def\@makefigurecaption#1#2{\vskip 0mm #1. #2\par}

\long\def\@maketablecaption#1#2{\hbox to \hsize{\parbox[t]{\hsize}
  {#1 \\ #2}}\vskip 0.3ex}

\def\fnum@figure{Figure \thefigure}
\def\figure{\let\@makecaption\@makefigurecaption \@float{figure}}
\@namedef{figure*}{\let\@makecaption\@makefigurecaption \@dblfloat{figure}}

\def\table{\let\@makecaption\@maketablecaption \@float{table}}
\@namedef{table*}{\let\@makecaption\@maketablecaption \@dblfloat{table}}

\textfloatsep=\floatsep           
\intextsep=\floatsep              

\long\def\@makefntext#1{\parindent 1em\noindent\hbox{${}^{\@thefnmark}$}#1}

\def\maketitle{\begingroup        
    \def\thefootnote{\fnsymbol{footnote}}%
    \newpage \global\@topnum\z@
    \@maketitle \@thanks
  \endgroup
  \let\maketitle\relax \let\@maketitle\relax
  \gdef\@thanks{}\let\thanks\relax
  \gdef\@address{}\gdef\@author{}\gdef\@title{}\let\address\relax}

\def\justify@on{\let\\=\@normalcr
  \leftskip\z@ \@rightskip\z@ \rightskip\@rightskip}

\newbox\fm@box                    

\def\@maketitle{
  \global\setbox\fm@box=\vbox\bgroup
    \vskip 8mm                    
    \raggedright                  
    \hyphenpenalty\@M             
    {\Large \@title \par}         
    \vskip\@bls                   
    {\normalsize                  
     \@author \par}               
    \vskip\@bls                   
    \@address                     
  \egroup
  \twocolumn[
    \unvbox\fm@box                
    \vskip\@bls                   
    \unvbox\abstract@box          
    \vskip 2pc]}                  

\newcounter{address}
\def\theaddress{\alph{address}}
\def\@makeadmark#1{\hbox{$^{\rm #1}$}}

\def\address#1{\addressmark\begingroup
  \xdef\@tempa{\theaddress}\let\\=\relax
  \def\protect{\noexpand\protect\noexpand}\xdef\@address{\@address
  \protect\addresstext{\@tempa}{#1}}\endgroup}
\def\@address{}

\def\addressmark{\stepcounter{address}%
  \xdef\@tempb{\theaddress}\@makeadmark{\@tempb}}

\def\addresstext#1#2{\leavevmode \begingroup
  \raggedright \hyphenpenalty\@M \@makeadmark{#1}#2\par \endgroup
  \vskip\@bls}

\newbox\abstract@box              

\def\abstract{%
  \global\setbox\abstract@box=\vbox\bgroup
  \small\rm
  \ignorespaces}
\def\endabstract{\par \egroup}

\def\thebibliography#1{\section*{REFERENCES}\list{\arabic{enumi}.}
  {\settowidth\labelwidth{#1.}\leftmargin=1.67em
   \labelsep\leftmargin \advance\labelsep-\labelwidth
   \itemsep\z@ \parsep\z@
   \usecounter{enumi}}\def\makelabel##1{\rlap{##1}\hss}%
   \def\newblock{\hskip 0.11em plus 0.33em minus -0.07em}
   \sloppy \clubpenalty=4000 \widowpenalty=4000 \sfcode`\.=1000\relax}

\newcount\@tempcntc
\def\@citex[#1]#2{\if@filesw\immediate\write\@auxout{\string\citation{#2}}\fi
  \@tempcnta\z@\@tempcntb\m@ne\def\@citea{}\@cite{\@for\@citeb:=#2\do
    {\@ifundefined
       {b@\@citeb}{\@citeo\@tempcntb\m@ne\@citea
        \def\@citea{,\penalty\@m\ }{\bf ?}\@warning
       {Citation `\@citeb' on page \thepage \space undefined}}%
    {\setbox\z@\hbox{\global\@tempcntc0\csname b@\@citeb\endcsname\relax}%
     \ifnum\@tempcntc=\z@ \@citeo\@tempcntb\m@ne
       \@citea\def\@citea{,\penalty\@m}
       \hbox{\csname b@\@citeb\endcsname}%
     \else
      \advance\@tempcntb\@ne
      \ifnum\@tempcntb=\@tempcntc
      \else\advance\@tempcntb\m@ne\@citeo
      \@tempcnta\@tempcntc\@tempcntb\@tempcntc\fi\fi}}\@citeo}{#1}}

\def\@citeo{\ifnum\@tempcnta>\@tempcntb\else\@citea
  \def\@citea{,\penalty\@m}%
  \ifnum\@tempcnta=\@tempcntb\the\@tempcnta\else
   {\advance\@tempcnta\@ne\ifnum\@tempcnta=\@tempcntb \else
\def\@citea{--}\fi
    \advance\@tempcnta\m@ne\the\@tempcnta\@citea\the\@tempcntb}\fi\fi}

\def\ps@crcplain{\let\@mkboth\@gobbletwo
     \def\@oddhead{\reset@font{\sl\rightmark}\hfil \rm\thepage}%
     \def\@evenhead{\reset@font\rm \thepage\hfil\sl\leftmark}%
     \let\@oddfoot\@empty
     \let\@evenfoot\@oddfoot}

\ps@crcplain                    

\def\beq{\begin{equation}}
\def\eeq{\end{equation}}
\let\dsp=\displaystyle
\def\O{{\cal O}}
\def\Dsl{ \Delta\!\!\! / \, }
\long\def \omit #1 {}

\catcode`\@=11
\def\@versim#1#2{\lower 2\p@\vbox{\baselineskip\z@skip\lineskip+1\p@
    \ialign{$\m@th#1\hfil##\hfil$\crcr#2\crcr\sim\crcr}}}
\def\gapp{\mathrel{\mathpalette\@versim>}}
\def\lapp{\mathrel{\mathpalette\@versim<}}

\def \reset@font {}

\catcode`\@=12

\newcommand{\ttbs}{\char'134}
\newcommand{\AmS}{{\protect\the\textfont2
  A\kern-.1667em\lower.5ex\hbox{M}\kern-.125emS}}

\title{The D234 action for light quarks }

\author{
M. Alford\thanks{Address after Sept 1995:
School of Natural Sciences, Institute for Advanced Study, Princeton, NJ 08540},
    T. Klassen\thanks{Address after Sept 1995:
	SCRI, Florida State University, Tallahassee, FL 32306},
 and P. Lepage \\ ~\\
        {Newman Laboratory of Nuclear Studies \\
                 Cornell University \\
                 Ithaca, NY 14853 \\
                 USA }
        }

\begin{document}

\begin{abstract}
We investigate a new light fermion action (the ``D234'' action), which
is accurate up to $\O(a^3)$ and tadpole-improved $\O(a \alpha_s)$ errors.
Using D234 with Symanzik- and tadpole-improved glue we find evidence
that continuum results for the quenched hadron spectrum (pion, rho and nucleon)
can be obtained on coarse lattices.
\end{abstract}

\kern -10ex
\maketitle

\section{INTRODUCTION}

It has recently become clear that improved actions \cite{sym,lw}
are a powerful and practical tool for obtaining continuum results from
the lattice. This has been demonstrated for the $\Upsilon$ with the
non-relativistic QCD action \cite{nrqcd}, and also for the static
quark potential and charmonium with improved glue
\cite{heavy}. In the latter case, continuum results were obtained
from lattices as coarse as $a=0.4$ fm.

At tree level, improved glue has $\O(a^4)$ errors, much smaller than the
$\O(a^2)$ errors of the Sheikholeslami-Wohlert (SW) action \cite{sw}
 for fermions,
which is the most improved light quark action used in simulations up
to now (see other contributions at this conference). In order to bring the
quark action up to the same level of accuracy as the glue,
we have started a program
of investigating improved light fermion actions.
We here report results for the ``D234'' action in the quenched
approximation. The D234 action uses a cubic
derivative term to improve the naive lattice Dirac operator.
It also contains
second and fourth order derivative terms (as well as the $\sigma \cdot F$
``clover'' term) with fine tuned coefficients to eliminate the ``spatial
doublers'' at large momentum and minimize the number of ``time doublers'' at
low momentum without introducing errors larger than $\O(a^3)$.

As with other improved actions, we find that tadpole
improvement \cite{tadp} of the coefficients is
essential in obtaining accurate results with the D234 action.
At tree level it can be implemented by simply replacing
each gauge link $U$ in the action
by $\tilde U = U/u_0$, where $(u_0)^4$ is the expectation value of the
plaquette.

\section{D234 ACTION}

The D234 fermionic operator used in the simulations described here is:
{ \setlength{\arraycolsep}{0em}
\begin{eqnarray}
&&\dsp m_0 + \sum_\mu
  \biggl( \gamma_\mu\Delta^{(1)}_\mu -
 ba^2 \gamma_\mu\Delta^{(1)}_\mu  \Delta^{(2)}_\mu
   \nonumber \\
&&- \frac{ra}{2} \Bigl( \Delta^{(2)}_\mu
 + \frac{1}{2} \sum_\nu \sigma_{\mu\nu} F_{\mu\nu} \Bigr)
 + ca^3 \Delta^{(2)}_\mu \Delta^{(2)}_\mu \biggr),
\end{eqnarray}
}
where $b=1/6, r=2/3, c=1/12$,
{ \setlength{\arraycolsep}{0.2em}
\begin{eqnarray}
 \Delta^{(1)}_\mu \psi(x) &=& {\dsp {1\over 2a}} 
   \Bigl( \tilde U_\mu(x)\psi(x+\mu) \nonumber \\
  && - \, \tilde U_{-\mu}(x)\psi(x-\mu) \Bigr), \\
\Delta^{(2)}_\mu \psi(x) &=& {\dsp {1\over a^2}} 
   \Bigl( \tilde U_\mu(x)\psi(x+\mu)  \nonumber \\
 && + \, \tilde U_{-\mu}(x)\psi(x-\mu) - 2 \psi(x) \Bigr),  
\end{eqnarray}
}
and for $F_{\mu\nu}$ we use the clover representation \cite{sw}.

\begin{figure}[htb]
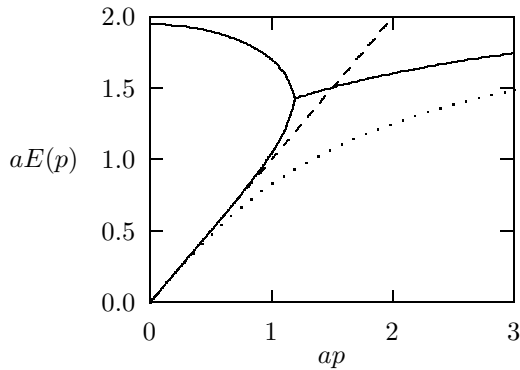

\setlength{\unitlength}{0.240900pt}
\ifx\plotpoint\undefined\newsavebox{\plotpoint}\fi
\sbox{\plotpoint}{\rule[-0.175pt]{0.350pt}{0.350pt}}%


\caption{\label{fig:disp} Massless dispersion relations for
free D234 (solid lines), Wilson or SW (dotted line) and
continuum (dashed line) fermions.
Momentum $p$ is along the (1,1,0) direction. Beyond the D234 branch point,
the real part of the two conjugate roots is shown.}
\end{figure}

Classically the D234 action has  $\O(a^3)$ errors.
One can show this by using a change of variable to relate it to
$\gamma_\mu\Delta^{(1)}_\mu (1 - \frac{a^2}{6} \Delta^{(2)}_\mu)$,
which differs at $\O(a^4)$ from the continuum Dirac operator.
For the rest of this paper we ignore the change of variable
since it does not affect spectral quantities.
\omit{
(note that
the $a^3$ error can be eliminated by suitably adjusting $b$ at $\O(m_0)$),
which can be shown by deriving it via a change of variable from the
fermionic operator
$\gamma_\mu\Delta^{(1)}_\mu (1 - \frac{a^2}{6} \Delta^{(2)}_\mu)$,
which has $\O(a^4)$ errors. }
At the 1-loop level there are additional $\O(a g^2)$ errors.
The coefficients $r$ and $c$ are chosen so that there is only one low
momentum doubler in the free dispersion relation,
as can be seen in fig.~\ref{fig:disp}.
For the glue we used the Symanzik- and tadpole-improved action
described in \cite{lw,heavy},
which has $O(a^4)$ and $O(a^2 g^2)$ errors.
\omit{We rely on
tadpole improvement to minimize the coefficients of the $\O(a g^2)$ and
$O(a^2 g^2)$ errors.
}

\section{PROCEDURE}

The results reported here were obtained using code written in C++ (with matrix
multiplications in FORTRAN for added speed), and run on Cornell's SP-2.
Gluon configurations were generated by the Metropolis algorithm,
with 10 hits per update and 20 $-$ 30 updates between configurations, which
suffices to decorrelate them on the volumes we used.
The quark propagator in each gauge background was found using
the BiCGStab2 algorithm, supplemented by a simple type of preconditioning.

We performed simulations at three couplings,
$\beta \! \equiv \! \beta_{\rm pl} \! = \! 6.8, 7.1, 7.4$,
corresponding to lattice spacings $a= 0.40, 0.33, 0.24$ fm, if one uses
charmonium  to set the scale \cite{heavy}.
We have results for seven quark masses at $\beta$=6.8 and 7.1, for
four at $\beta$=7.4, with
$m_\rho/m_\pi$ ratios in the range between 1.2 to 2.0.
For the smaller ratios the spatial size of our lattices is about 2.0 fm,
increasing to 2.6 $-$ 3.2 fm for larger ratios, except for $\beta$=7.4,
where at present  our largest lattices are 2.2 fm.
In lattice units our spatial volumes are in the  range $5^3$ to $9^3$.
The time extent of our lattices is
always as least twice the spatial one. Since the cost of generating gluon
configurations is negligible compared to that of the inversion, we did not
save configurations but generated  independent ones for each quark mass.
We typically produced between 200 and 600 propagators per quark mass, the
smaller numbers corresponding to $\beta$=7.4.

We calculated the time slice propagators for the relativistic pion,
rho and nucleon operators, as well as for the non-relativistic nucleon,
and the $J_z$=$1/2$ and $3/2$ delta operators.
 Furthermore, we obtained
the pion and rho with various momenta. 	In all cases we suitably
averaged over spin so that (in the infinite statistics limit) the time
slice propagators for large times $t$ are $\propto \cosh m (\frac{T}{2}-t)$,
where T is the time extension of the lattice.

We used a covariant form of gaussian smearing for the
quark fields at the source and sink, employing the operator
$ \left( 1 + {\epsilon}\sum_i\Delta^{(2)}_i \right)^N $.
On our coarse lattices it seems that for a given $\beta$ one can use the same,
near-optimal (in terms of cost/error) value of $\epsilon N$
for all quark masses and hadrons.
Initially we used
(correlated) 2x2 matrix fits of all four combinations of
sink and source local and smeared propagators
(LL, SL, LS, SS) to ground and excited state cosh's.
After determining the values of $\epsilon N$ where the
SS propagator has only a very small excited state contribution,
we found it more cost efficient to generate only the source smeared
LS and SS propagators and fit them to a single cosh, either separately
or in a combined fit.
We used a reliable, if perhaps somewhat conservative method
to combine effective mass estimates from fits with different $t_{\rm min}$,
instead of picking the fit with the best ``goodness'' by some
criterion. This is relevant
in cases where fits from different $t_{\rm min}$ appear to be
nearly equally ``good''.

\section{RESULTS}

We first investigated finite volume effects. We find that for
$m_\rho/m_\pi \!\approx \! 2$, infinite volume results are reached within
our statistical errors for the pion and rho
at spatial sizes of around $2.0$ fm, for the nucleon around
$2.5$ fm.  For smaller $m_\rho/m_\pi$
ratios, somewhat smaller volumes suffice. All our results presented below
should therefore have no significant finite volume errors, except perhaps
the nucleon at $\beta$=7.4 (cf.~below). The delta might require even bigger
volumes than the nucleon, and we will not present results for it, except
to remark that the good agreement we find between the masses of the
$J_z$=$1/2$ and $3/2$ delta states, which are as large as $a m_\Delta
\! \approx \! 3$, provide good evidence for the rotational invariance of the
D234 action on coarse lattices.

Further such evidence is furnished by the pion and rho states with
momenta. Defining  a ``speed of light'' $c$ via
\beq
\label{eq:csq}
c^2 p^2 = E(p)^2 - E(0)^2 ~,
\eeq
we  compare in table~\ref{tab:csq}
results from the D234 and SW actions.
We see that rotational invariance, i.e.~$c^2 \approx 1$, is restored
already on the coarsest lattice for D234, in contrast to SW. This conclusion
is even more striking for the higher momenta we investigated.
We have also checked that $c$ deteriorates dramatically without tadpole
improvement in the D234 action, even though it is still
better than for SW with tadpole improvement (with improved glue in both cases).

\def\st{\rule[-1.5ex]{0em}{4ex}} 
\begin{table}[tb]
\setlength{\tabcolsep}{0.7pc}
\newlength{\digitwidth} \settowidth{\digitwidth}{\rm 0}
\caption{
``Speed of light'' squared ($c^2$ in eq.~\protect\ref{eq:csq})
for mesons with momentum $p \! = \! 2\pi/aL$, $aL\! \approx \! 2.0$ fm
at $m_\pi/m_\rho \approx 0.70$.}
\label{tab:csq}
\begin{tabular*}{75mm}{lllll}
\hline
& \makebox[1mm][l]{D234 Action} &&  \makebox[1mm][l]{SW Action} \\
\st $\beta$ & $\pi$ & $\rho$ & $\pi$ & $\rho$ \\
\hline
6.8  &  0.95(2)  & 0.93(3) & 0.63(2) & 0.48(3) \\
7.1  &  0.94(3)  & 0.96(5) & 0.74(3) & 0.55(4) \\
7.4  &  0.99(4)  & 1.00(6) & --- & --- \\
\hline
\end{tabular*}
\end{table}

\begin{figure}[htb]
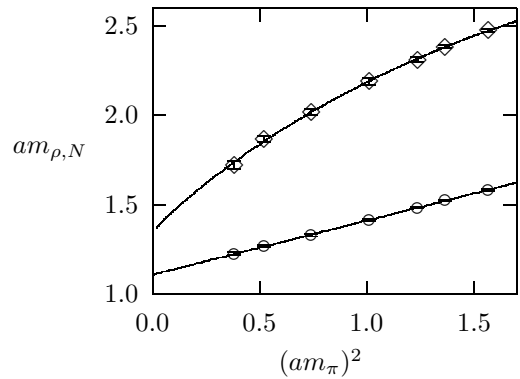

\setlength{\unitlength}{0.240900pt}
\ifx\plotpoint\undefined\newsavebox{\plotpoint}\fi


\caption{\label{fig:rhoN} Masses of the rho (circles) and nucleon
(diamonds) and their chiral fits at $\beta \! = \! 7.1$.}
\end{figure}

We now describe our results for the hadron masses. The data are shown in
table~\ref{tab:masses}.
For $m_\rho$ we can obtain perfect linear fits to $V_0 + V_2 m_\pi^2$
up to the largest masses considered, as shown in table~\ref{tab:fitparsrho},
and in fig.~\ref{fig:rhoN} for $\beta \! = \! 7.1$.
Lacock and Michael \cite{j} proposed to use the dimensionless parameter
$J \!= \! m_V d m_V/d m_P^2$ at $m_V/m_P \!=\! m_{K^\star}/m_K \!= \!1.8$ to
characterize the vector versus pseudo-scalar mass relation and compare lattice
studies to nature. They estimate $J\!=\!0.48(2)$ for the real world
--- note that $J\!=\!0.5$ if $m_V^2 - m_P^2$ were exactly constant ---
 and $J\!=\!0.37(2)(4)$
as the ``world average'' from quenched simulations. Our estimate of $J$ from
chiral fits is shown in
table~\ref{tab:resl}.
Other estimates in these proceedings are
also in excellent agreement with ours, but do not scale on similarly
coarse lattices.

\begin{table}[hbt]
\setlength{\tabcolsep}{0.5pc}
\catcode`?=\active \def?{\kern\digitwidth}
\caption{Light hadron masses from the D234 action}
\label{tab:masses}
\begin{tabular*}{75mm}{lllll}
\hline
\st $\beta$ & $a m_0$  & $a m_\pi$ & $a m_\rho$ & $a m_N$ \\
\hline
6.8  & -0.15  &  0.700(4)   & 1.375(10)  & 2.047(24)\\
     & -0.08  &  0.911(4)   & 1.466(8)   & 2.270(17)\\
     & -0.05  &  0.980(3)   & 1.497(6)   & 2.331(16)\\
     & ~0     &  1.092(3)   & 1.566(6)   & 2.464(15)\\
     & ~0.05  &  1.206(3)   & 1.638(5)   & 2.598(12)\\
     & ~0.1   &  1.298(2)   & 1.698(5)   & 2.695(10)\\
     & ~0.2   &  1.480(3)   & 1.838(5)   & 2.951(14)\\
\hline
7.1  & -0.14  &  0.616(4)   & 1.226(8)   & 1.720(21)\\
     & -0.11  &  0.721(3)   & 1.266(5)   & 1.868(17)\\
     & -0.06  &  0.861(3)   & 1.327(7)   & 2.018(16)\\
     & ~0     &  1.005(3)   & 1.413(7)   & 2.189(19)\\
     & ~0.05  &  1.112(2)   & 1.483(5)   & 2.311(15)\\
     & ~0.08  &  1.168(2)   & 1.524(4)   & 2.384(10)\\
     & ~0.12  &  1.251(2)   & 1.581(6)   & 2.474(10)\\
\hline
7.4  & -0.05  &  0.662(4)   & 1.055(9)   & 1.560(15)\\
     & ~0     &  0.806(4)   & 1.147(7)   & 1.739(10)\\
     & ~0.05  &  0.921(5)   & 1.224(9)   & 1.869(14)\\
     & ~0.1   &  1.031(3)   & 1.304(4)   & 2.010(9)\\
\hline
\end{tabular*}
\end{table}

\begin{table}[hbt]
\setlength{\tabcolsep}{0.6pc}
\catcode`?=\active \def?{\kern\digitwidth}
\caption{Chiral fit parameters (in lattice units) of the rho mass}
\label{tab:fitparsrho}
\begin{tabular*}{75mm}{llllll}
\hline
\st $\beta$  & $ V_0$  & $ V_2$  & $N_{\rm DF}$  &  $Q$\\
\hline
6.8  & 1.238(7)   & 0.274(5)      & 5   & 0.98\\
7.1  & 1.108(6)   & 0.303(5)      & 5   & 0.94\\
7.4  & 0.889(13)  & 0.390(14)     & 2   & 0.67\\
\hline
\end{tabular*}
\end{table}

\begin{table}[hbt]
\setlength{\tabcolsep}{0.35pc}
\catcode`?=\active \def?{\kern\digitwidth}
\caption{Chiral fit parameters (in lattice units) of the nucleon mass}
\label{tab:fitparsN}
\begin{tabular*}{75mm}{llllll}
\hline
\st $\beta$ ~~& $N_0$ & $ N_2$ & $ N_3$  & $N_{\rm DF}$ & $Q$ \\
\hline
6.8  & 1.66(8) & 0.99(21) & -0.29(13) & 3 & 0.91\\
7.1  & 1.35(5) & 1.30(18) & -0.47(12) & 4 & 0.80\\
7.4  & 1.09(11)& 1.44(46) & -0.56(35) & 1 & 0.55\\
\hline
\end{tabular*}
\end{table}

We now turn to the nucleon. The masses
extracted from the relativistic and non-relativistic operators are found to
agree well within statistical errors.
To obtain good chiral fits it is crucial, in general, to include a cubic
term, i.e.~fit $m_N$ to $N_0 + N_2 m_\pi^2 + N_3 m_\pi^3$,
cf.~fig~\ref{fig:rhoN}. Our results are shown in table~\ref{tab:fitparsN}.
For $\beta$=7.4 we can not completely exclude finite volume errors in the
chirally extrapolated $m_N$,
but they are almost certainly smaller than the large statistical errors we have
in this case.
\omit{
{}From tables~\ref{tab:fitparsrho} and~\ref{tab:fitparsN} one can obtain the
chirally extrapolated mass ratio $m_N/m_\rho \! = \! 1.34(6), 1.22(5),
1.23(12)$
(=1.22 in nature) at $\beta \! = \! 6.8, 7.1, 7.4$, respectively.
}
 In table~\ref{tab:resl}
we show our estimates of $m_N/m_\rho$ (=1.22 in nature) in the chiral limit.

So far the results described for the D234 action have been spectacular.
There is however one problem, which is revealed in table~\ref{tab:resl}:
the scale determined from the chirally extrapolated $\rho$ mass (set to
$770$ MeV) does not agree with that from the charmonium 1S$-$1P
splitting ($a_\psi$),
which is known to scale quite well~\cite{heavy}.
In view of the excellent properties of $c^2$ and $J$ we interpret this as
indicating a small additive shift of all masses (vanishing, of course, as
$a \! \rightarrow \! 0$)
that depends, it seems, only rather weakly on the momentum
or type of the particle, at least for mesons.
We are investigating whether the doublers are
responsible for this shift.
There are various strategies to
``push away'' the doublers, and we are hopeful that a completely
satisfactory fermionic action for coarse lattices can be found.

\def\fm{\hbox{fm}}

\begin{table}[bt]
\setlength{\tabcolsep}{0.3pc}
\catcode`?=\active \def?{\kern\digitwidth}
\caption{
Selected quantities relevant for scaling.
$m_N/m_\rho$ is extrapolated to zero
pion mass, $a_\psi$ is from~\protect\cite{heavy}.}
\label{tab:resl}
\begin{tabular*}{75mm}{lllll}
\hline
\st $\beta$ & $J$  & $m_N/m_\rho$ & $a^{-1}_\psi$ (MeV) & $a_\psi/ a_\rho$ \\
\hline
6.8  & 0.384(8)   & 1.34(6) & 497(4) & 1.251(12) \\
7.1  & 0.381(8)   & 1.22(5) & 606(5) & 1.147(11) \\
7.4  & 0.395(18)  & 1.23(12) & 821(10) & 1.055(20) \\
\hline
\end{tabular*}
\end{table}

{\em Acknowledgements.\/}
This research was conducted using the resources of the Cornell Theory
Center, which receives major funding from the
NSF
and New York State, with additional support from
ARPA,
 the National Center for
Research Resources at the
NIH, IBM
 and other members of the center's Corporate Research
Institute.

\end{document}